\documentclass[aps,prc,preprint,groupedaddress,showpacs,showkeys,nofootinbib]{revtex4}
\usepackage{epsfig,psfig,float,amssymb,latexsym}
\newcommand{\be}{\begin{equation}}
\newcommand{\ee}{\end{equation}}
\newcommand{\ba}{\begin{eqnarray}}
\newcommand{\ea}{\end{eqnarray}}

\begin{document}

\preprint{IFIC$-$04$-$0302}
\preprint{FTUV$-$04$-$0302}

\title{Kaon-antikaon nuclear optical potentials \\
and the $\kappa$ meson in the nuclear medium} 

\author{D. Cabrera}
\email[]{Daniel.Cabrera@ific.uv.es}
\affiliation{Departamento de F\'{\i}sica Te\'orica and IFIC, 
Centro Mixto Universidad de Valencia-CSIC, 
Institutos de Investigaci\'on de Paterna, Apdo. correos 22085, 
46071, Valencia, Spain}
\author{M.J. Vicente Vacas}
\email[]{Manuel.J.Vicente@ific.uv.es}
\affiliation{Departamento de F\'{\i}sica Te\'orica and IFIC, 
Centro Mixto Universidad de Valencia-CSIC, 
Institutos de Investigaci\'on de Paterna, Apdo. correos 22085, 
46071, Valencia, Spain}

\date{\today}

\begin{abstract}
We study the properties of the $\kappa$ meson in the nuclear medium, starting
from a chiral unitary model of $S-$wave, I=1/2 $K\pi$ scattering, which
describes the elastic $K\pi$ phase shifts and generates a pole in the
amplitude. Medium effects are considered by including pion and kaon
selfenergies. We explore the changes in the $\kappa$ pole position and in the
$K\pi$ scattering amplitude at finite density, with two different models for the
kaon selfenergy.

\end{abstract}

\pacs{13.25.-k; 14.40.Ev; 24.10.Eq}
\keywords{kappa meson, medium effects, meson selfenergy, chiral unitary approach}

\maketitle

\section{\label{s1}Introduction}

   Kaon and antikaon properties in the nuclear medium are quite different.
Whereas the kaons selfenergy is weak, repulsive and can be well accounted for 
by a $t\rho$ approximation, the situation is much more complicated for the 
antikaons, among other reasons because of the presence of the  $\Lambda (1405)$
resonance just  below the $\bar{K} N$ threshold. The antikaon selfenergy is
particularly interesting because if it is sufficiently attractive it could lead
to the existence of a $\bar{K}$ condensed phase in the neutron stars, as it was
pointed out in \cite{Kaplan:1986yq}, (for a recent discussion see
\cite{Ramos:2000dq}). 

There is already clear experimental evidence of strong medium effects that
deviate  from simple low density theorem predictions on the strong enhancement
of the $K^-$ production in heavy-ion collisions in the KaoS experiments  at
GSI \cite{Laue:1999yv}. Although this result could imply a very attractive 
potential it is difficult to extrapolate it to the high density and low
temperature relevant for the neutron stars. 

Also strong medium effects have been found in kaonic atoms. However, there is
a strong controversy on the theoretical interpretation of these data, with 
widely different potentials available in the literature which provide a
reasonable agreement to the data. They could be classified in two
basic kinds, some
phenomenological potentials, usually deeply attractive, 
\cite{Sibirtsev:1998vz,Friedman:cu,Friedman:hx,Friedman:1998xa}
and other chirally motivated potentials 
\cite{Kaiser:1995eg,Waas:xh,Waas:fy,Lutz:1997wt,Oset:1997it,Ramos:1999ku,Hirenzaki:2000da,Baca:2000ic,Cieply:2001yg},
which after the need 
of a selfconsistent treatment of the kaon and the $\Lambda (1405)$ in the
 medium was shown \cite{Lutz:1997wt}, are clearly less attractive.

Some other observables that could be sensitive to the kaonic nuclear potential
have been suggested, like the study of the $\phi$ mass in nuclear reactions.
However, due to the cancellation between attraction and repulsion for the
antikaon and kaon respectively little if some effect is expected
\cite{Cabrera:2002hc}.

On the other hand some interesting developments have taken place in the
study of the light scalar mesons, in particular the $\sigma$ and the 
$\kappa$. The $\sigma$ meson now appears to be quite
well established both theoretically and experimentally \cite{Hagiwara:fs}
although some discussion on its nature still remains open. In particular,
unitarized models consistent with chiral constraints describe the  $\pi\pi$
phase shifts and clearly predict a pole at masses  of around 470 MeV
\cite{Oller:1998zr,Colangelo:2001df}. Some works using this kind of chiral
models have also studied the isospin $I=1/2$ channel and find a very wide
$\kappa$ meson at masses around 800 MeV. A compilation of results both
experimental and theoretical can be found in 
\cite{vanBeveren:2002vw,Ochs:2003hb,Bugg:kj}.

In the unitarized chiral models the $\sigma$  and $\kappa$ mesons can be
understood as $\pi\pi$ and $\pi K$ resonances and this has important
consequences for their interaction with the nuclear medium, as their selfenergy
will be directly related to that of the $\pi$ and $K$ mesons. The $\sigma$ 
medium properties have been studied using different approaches finding always a
strong reduction of its mass and a much narrower width at  high baryonic
densities \cite{Rapp:1995ir,Chiang:1997di,Hatsuda:kd}. There are some
experimental signals that strongly suggest that this is indeed the case. A quite
strong enhancement of the  $\pi\pi$ invariant mass spectrum at the low masses
predicted for the in-medium $\sigma$  has been found in both $(\pi,\pi\pi)$ 
\cite{Bonutti:zw,Bonutti:ij,Rapp:1998fx,VicenteVacas:1999xx} and $(\gamma,\pi\pi)$ reactions
\cite{Messchendorp:2002au,Roca:2002vd,Muhlich:2004zj}. 
In the unitarized chiral models the mass reduction of the $\sigma$, and its
narrowing in the nuclear medium  are mainly
produced by the well known attractive p-wave pion-nucleus optical potential.
A similar, although richer in complexity, situation could occur for the 
$\kappa$ meson. Given the clearly different interaction of the kaons and
antikaons with the medium one could expect a splitting of the masses of $\kappa$
and anti-$\kappa$. The difference would be sensitive  to the different optical
potential suffered by kaons and antikaons. Furthermore, as for the $\sigma$
case, the strong attraction over the pion could lead to a common mass reduction
and a narrower width.

Our purpose in this paper is to investigate this possibility. We start by
presenting a simple model that predicts a $\kappa$ pole and is consistent with
the meson-meson phenomenology at low and intermediate energies in the next
section. Next, we incorporate the nuclear medium effects using for that two different
potentials in order to  study the sensitivity of the observables to these
potentials.

\section{\label{s2}$K \pi$ scattering in a chiral unitary approach}
We briefly revise in this chapter the model of $K \pi$ scattering, which is
based on the chiral unitary approach to meson meson scattering developed in
\cite{Oller:1997ti}. In that work, a good agreement with experimental
phase shifts for the $S-$wave meson meson scattering in the $I=0,1$ channels was
found.
We solve the Bethe-Salpeter (BS) equation, namely $T=V+VGT$, in
which the kernel $V$ is taken as the $I=1/2$ $K\pi$ tree level amplitude from
the lowest order $\chi$PT Lagrangian. After projecting onto the
$S-$wave, this amplitude reads
\be
\label{Vkpitree}
V=\frac{1}{4f^2} \left[ -\frac{5}{2} s + m_{\pi}^2 + m_{K}^2
+ \frac{3}{2} \frac{(m_{K}^2-m_{\pi}^2)^2}{s} \right] \,\,.
\ee
In Eq. (\ref{Vkpitree}), $s$ is the Mandelstam variable, $m_{\pi}$ and $m_{K}$
are the pion and kaon masses, respectively, and $f$ is the meson decay constant,
which we take to be $f^2 = (100 \,\textrm{MeV})^2 \simeq f_{\pi} f_{K}$ as in
Ref. \cite{Oller:1998zr}.

\begin{figure}
\includegraphics[width=0.7\textwidth]{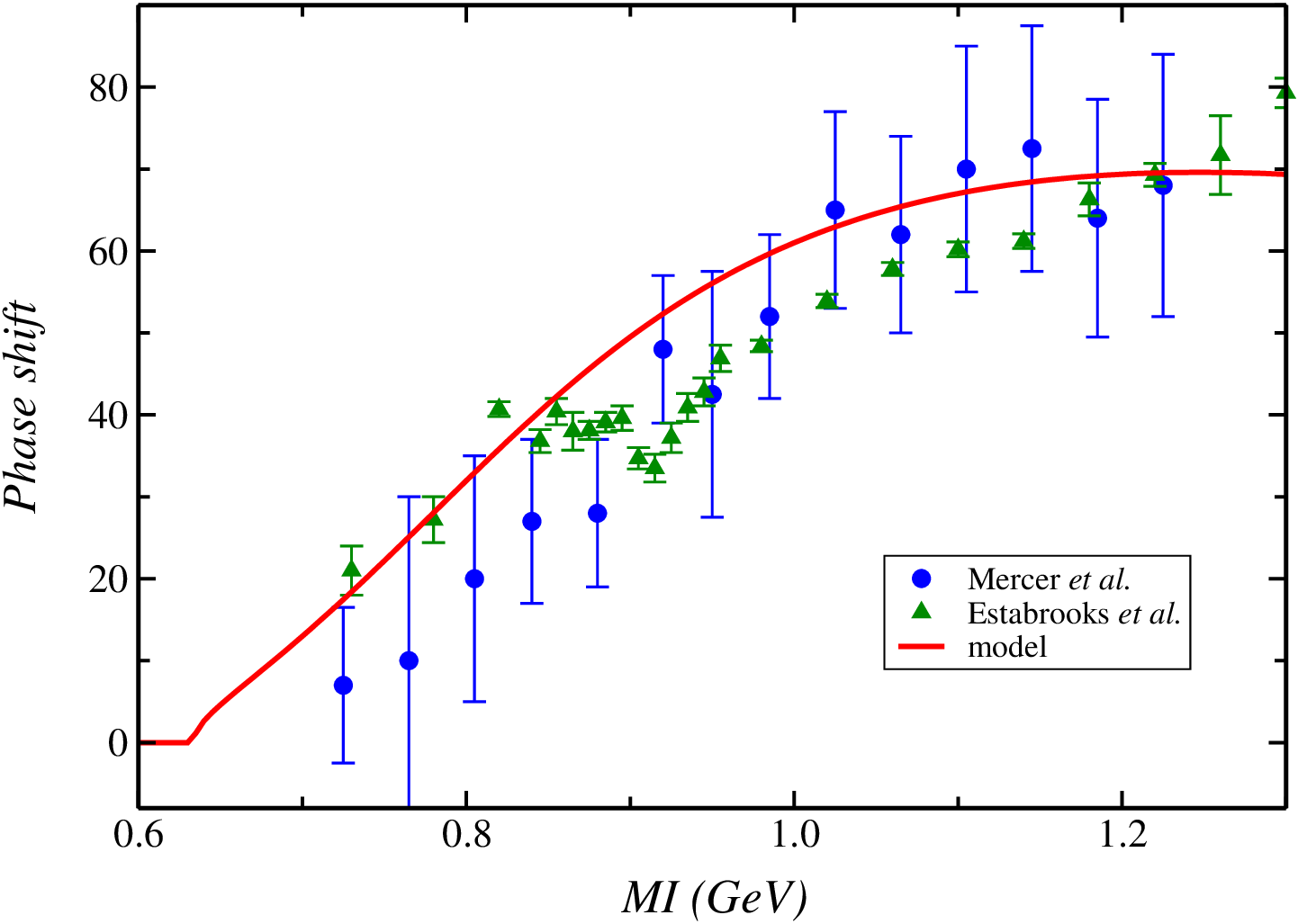}
\caption{\label{phaseshifts} (Color online) $K\pi \to K\pi$ phase shifts in the $I=1/2$
channel. The solid line is the result of the model described in the text. Data
sets are taken from Refs. \cite{Mercer:kn} (dots) and
\cite{Estabrooks:1977xe} (triangles).}
\end{figure}

Despite the integral character of the BS equation, it was found in \cite{Oller:1997ti}
that the $V$ amplitude factorizes on-shell out of the integral in the $VGT$
term, and so does $T$. Thus $T$ can be algebraically solved in terms of $V$ and
the $G$ function, which contains the integral of the two meson propagators,
\be
\label{Gfree}
G(\sqrt{s}) = \frac{1}{4\pi^2}\int_0^{q_{max}} dq \,
\frac{q^2}{\omega_{\pi}(q) \omega_{K}(q)} \,
\frac{\omega_{\pi}(q) + \omega_{K}(q)}
{s-(\omega_{\pi}(q)+\omega_{K}(q))^2+i\epsilon} \,\, ,
\ee
with $\omega_{\pi}(q)=\sqrt{q^2+m_{\pi}^2}$ and 
$\omega_{K}(q)=\sqrt{q^2+m_{K}^2}$. In Eq. (\ref{Gfree}) we regularize the $G$
function with a cut-off in the momentum of the mesons in the loop. A value of
$q_{max}=850$ MeV is used, what leads to a satisfactory fit to the $I=1/2$
$K\pi$ phase shifts, as shown in Fig. \ref{phaseshifts}. The scattering
amplitude in this channel exhibits a pole in the second Riemann sheet (2ndRS) of
the complex energy plane, which we identify with the $\kappa$ meson. The pole
position that we get is $M_{\kappa} + i\, \Gamma_{\kappa} /2 = 825 + 460\,i$ (MeV).
We have not considered in this calculation the contribution of the $K \eta$
intermediate state. Its inclusion leads to a coupled channel calculation, as
done in \cite{Oller:1997ti} for $\pi\pi$ and $\bar{K} K$ channels. However, it was
found in \cite{Oller:1998hw} that this channel barely mixes with the $K\pi$ channel.
We have also checked that accounting for it barely modifies the phase shifts in
the region beyond 1 GeV, and produces no visible effect at lower energies. Since
we are mainly interested in the medium effects on the $K\pi$ channel, we shall
ignore the $K\eta$ contribution from now on.

\section{\label{s3}$K \pi$ scattering in the nuclear medium}
We consider medium corrections to the $K\pi$ scattering amplitude by dressing
the meson propagators with appropriate self-energies regarding the interactions
of the mesons with the surrounding nucleons of the nuclear medium. These
self-energies have been worked out elsewhere and we include here a brief
description. Then we explain our method to search for the $\kappa$ pole at
finite baryon density.

\subsection{Pion self-energy in the medium}
The pion self-energy is driven by the excitation, in $P-$wave, of $ph$ and
$\Delta h$ pairs. A derivation of the $P-$wave pion selfenergy from the $\pi N$
scattering amplitude can be found in Ref. \cite{Ericson:gk}, chapter 5.
Additionally, a resummation of short range correlation terms in
terms of the Landau-Migdal parameter $g'$ is done. The final expression reads
\begin{equation}
\Pi_{\pi}(q^0,\vec{q};\rho) = {\cal F}\,^2(\vec{q}\,^2) \, \vec{q}\,^2 
\frac{(\frac{D+F}{2 f})^2 \, U(q^0,\vec{q};\rho)}{1-(\frac{D+F}{2 f})^2 \,
g' U(q^0,\vec{q};\rho)} \,\,,
\end{equation}
where $(D+F)/ (2 f)$ is the $\pi NN$ coupling from the lowest order chiral
Lagrangian, $U=U_N+U_{\Delta}$ is the Lindhard function for the $ph$ and
$\Delta h$ excitations and ${\cal F}(\vec{q}\,^2)$ is a monopolar form factor. We follow 
the notation of \cite{Oset:1990ey,Cabrera:2002hc}.

\subsection{Kaon and antikaon self-energies in the medium}
We shall consider here two different potentials for the kaons, in order to  test
the sensitivity of the results to the models and their ability to discriminate
between them. The first model is a chirally motivated potential
\cite{Ramos:1999ku} which provides a weak attraction for the antikaons. We shall
refer to it as 'model A'. The second one is a phenomenological potential
described in \cite{Sibirtsev:1998vz}, leading to a very strong attraction for
the antikaons. We shall call it 'model B'.

The $KN$ interaction is smooth at low energies, and both models use a $t\rho$
approximation. The kaon self-energy is given
by
\begin{equation}
\label{Kself}
\Pi_{K} (\rho)= C \, m_K^2 \, \rho / \rho_0 \,\, ,
\end{equation}
where $\rho$ is the nuclear density, $\rho_0$ stands for the normal nuclear
density and $C$ takes the value $0.13$ for model A \cite{Oset:2001eg} and $0.114$ for model
B \cite{Sibirtsev:1998vz}.

The $\bar{K}N$ interaction, however, is much richer at low energies. Model A
starts from \cite{Oset:1998it}, where the $S-$wave $\bar{K} N$ scattering was
studied in a chiral unitary model in coupled channels, leading to a 
successful description of many scattering observables, namely, threshold ratios
of $K^- p$ to several inelastic channels; $K^- p$ and $K^- n$ scattering
lengths; and $K^- p$ cross sections in the elastic and inelastic channels
($\bar{K}^0 n$, $\pi^0 \Lambda$, $\pi^{\pm} \Sigma^{\mp}$, $\pi^0 \Sigma^0$). 
Medium effects
were considered in \cite{Ramos:1999ku} to obtain an effective $\bar{K}N$
interaction in nuclear matter, from which the $S-$wave $\bar{K}$ self-energy
was obtained in a selfconsistent way. 
Finally, the $P-$wave contribution to the $\bar{K}$ self-energy,
arising from the excitation of $Yh$ pairs ($Y=\Lambda, \Sigma,
\Sigma^*(1385)$) was also included. Full details can be
found in \cite{Ramos:1999ku,Cabrera:2002hc}.
Model B, based on a dispersive
calculation of the kaon potentials, which uses as input the $K^{\pm} N$
scattering amplitudes,
finds a strong $\bar{K}$ potential that
drops to $-200$ MeV at normal nuclear density and zero momentum, and shows a
quite strong momentum dependence. We use the parametrization given in
\cite{Sibirtsev:1998vz},
\begin{equation}
\label{barKGiessen}
\Pi_{\bar{K}}^B (q,\rho) = - \lbrack \, 0.233 + 0.563 \,
\exp (-1.242 \, q / m_K) \, \rbrack \, m_K^2 \, \rho / \rho_0 \,\,.
\end{equation}

These two potentials have been chosen as representative examples of the
kaon-nucleus potentials currently discussed in the literature. Some 'deep'
potentials (around $200$ MeV depth at zero $\bar{K}$ momentum) are found in
phenomenological analysis like \cite{Friedman:cu,Friedman:hx} and also in the
dispersive calculation of model B \cite{Sibirtsev:1998vz} after some
approximations. However, selfconsistent derivations of the $\bar{K}$ optical
potential starting from a good description of $\bar{K} N$ scattering produce
'shallow' potentials ($\simeq 40$ to $60$ MeV depth) like model A \cite{Ramos:1999ku}
and the potential obtained in \cite{Lutz:1997wt}. Both kinds of potentials produce a good agreement with
kaonic atom data, and the depth of the optical potential cannot be resolved by
analysing these data as has been shown in \cite{Cieply:2001yg}.

Higher order effects in density in the kaon selfenergy, like those induced by
the short range correlations in the pion selfenergy have been studied in
\cite{Waas:1996tw,Kolomeitsev:2002pg}. The effects have been found to be very
small at the densities considered in this work, and we shall neglect any
contribution from this source.

\subsection{Search for the $\kappa$ pole}
To look for the $\kappa$ pole in the medium we follow the procedure described in
\cite{VicenteVacas:2002se} for the $\sigma$ meson in $\pi\pi$ scattering, conveniently adapted
to the present problem.
We need to evaluate the $K\pi$ scattering amplitude, $T$, in the 2ndRS of the
complex energy ($\sqrt{s}$) plane. The analytical structure of $T$ is driven by
the $G(\sqrt{s})$ function, which has a cut on the positive energy real axis.
The rest of the functions in the BS equation are analytical and single-valued.
To calculate $G(\sqrt{s})$ at finite baryon density, we use the following
spectral (Lehmann) representations of the meson propagators,
\ba
\label{dressed_prop}
D_{\pi} (q^0,\vec{q};\rho) &=&
\int_0^{\infty} d\omega \,2 \omega \,
\frac{S_{\pi}(\omega,\vec{q};\rho)}{(q^0)^2-\omega^2+i\epsilon}
\nonumber \\
D_{\bar{K} (K)} (q^0,\vec{q};\rho) &=&
\int_0^{\infty} d\omega \bigg ( \frac{S_{\bar{K} (K)}
(\omega,\vec{q};\rho)}{q^0-\omega+i\eta} - \frac{S_{K (\bar{K})}
(\omega,\vec{q};\rho)}{q^0+\omega-i\eta} \bigg ) \,\, ,
\ea
where $S_{\pi}$ and $S_{\bar{K}(K)}$ are the spectral functions of the pion 
and the antikaon (kaon), respectively, that are directly related to the
selfenergies.
After some basic manipulations, $G(\sqrt{s})$ can be written as
\be
\label{Gmed}
G(\sqrt{s}) = \frac{1}{2 \pi} \int_0^{\infty} dW \left[
\frac{1}{\sqrt{s}-W+i\eta} F_P (W)
- \frac{1}{\sqrt{s}+W} F_{NP} (W) \right] \,\, ,
\ee
and $F_P (W)$, $F_{NP} (W)$ (accompanying the 'pole' and 'non-pole' terms,
respectively) are
real, positively-defined functions independent of $\sqrt{s}$. 
For $\bar{K}\pi$ scattering they are given by
\ba
\label{FW}
F_P(W) = \int \frac{d^3q}{(2\pi)^3} \int_{-W}^W du \, \pi \,
S_{\pi} (\frac{W-u}{2},\vec{q};\rho)\, S_{\bar{K}} (\frac{W+u}{2},\vec{q};\rho)
\nonumber \\
F_{NP}(W) = \int \frac{d^3q}{(2\pi)^3} \int_{-W}^W du \, \pi \,
S_{\pi} (\frac{W-u}{2},\vec{q};\rho)\, S_{K} (\frac{W+u}{2},\vec{q};\rho) \,\,.
\ea
For $K \pi$ scattering $S_{\bar{K}}$ in Eq. (\ref{FW}) has to be replaced by
$S_{K}$, and vice versa.
The cut-off
in the meson momentum is included in these functions (not explicitly shown). As
an example, in vacuum $F_P=F_{NP} \equiv F_{K\pi}$ factorizes out of the
brackets in Eq. (\ref{Gmed}) and reads
\be
\label{Fvac}
F_{K\pi}(W) = \frac{1}{4\pi} \frac{q(W)}{W} \, \theta (W-(m_{\pi}+m_K)) \,\, ,
\ee
with $q(W)=\lambda^{1/2}(W^2,m_{\pi}^2,m_K^2) / 2W$ and $\lambda$ the K\"allen
function.

The first term in Eq. (\ref{Gmed}) is responsible for the cut in the real axis. It
is easily shown that the $G$ function in the 2ndRS can be written as
\be
\label{GmedII}
G^{2ndRS}(\sqrt{s}) = \frac{1}{2 \pi} \int_0^{\infty} dW \left[
\frac{1}{\sqrt{s}-W+i\eta} F_P (W)
- \frac{1}{\sqrt{s}+W} F_{NP} (W) \right]
+ i \, F_P (\textrm{Re}\sqrt{s}) \,\, .
\ee

\section{\label{s4}Results and discussion}
We have calculated the $K\pi$ and $\bar{K}\pi$ scattering amplitudes and found
the $\kappa$ pole position for several nuclear densities from $\rho=0$ to
$\rho=1.5\,\rho_0$. To discuss the results we shall distinguish between the
following three cases: (1) free pion, in-medium kaons; (2) in-medium pion, free
kaons; and (3) in-medium pions and kaons.

\begin{figure}
\includegraphics[width=0.7\textwidth]{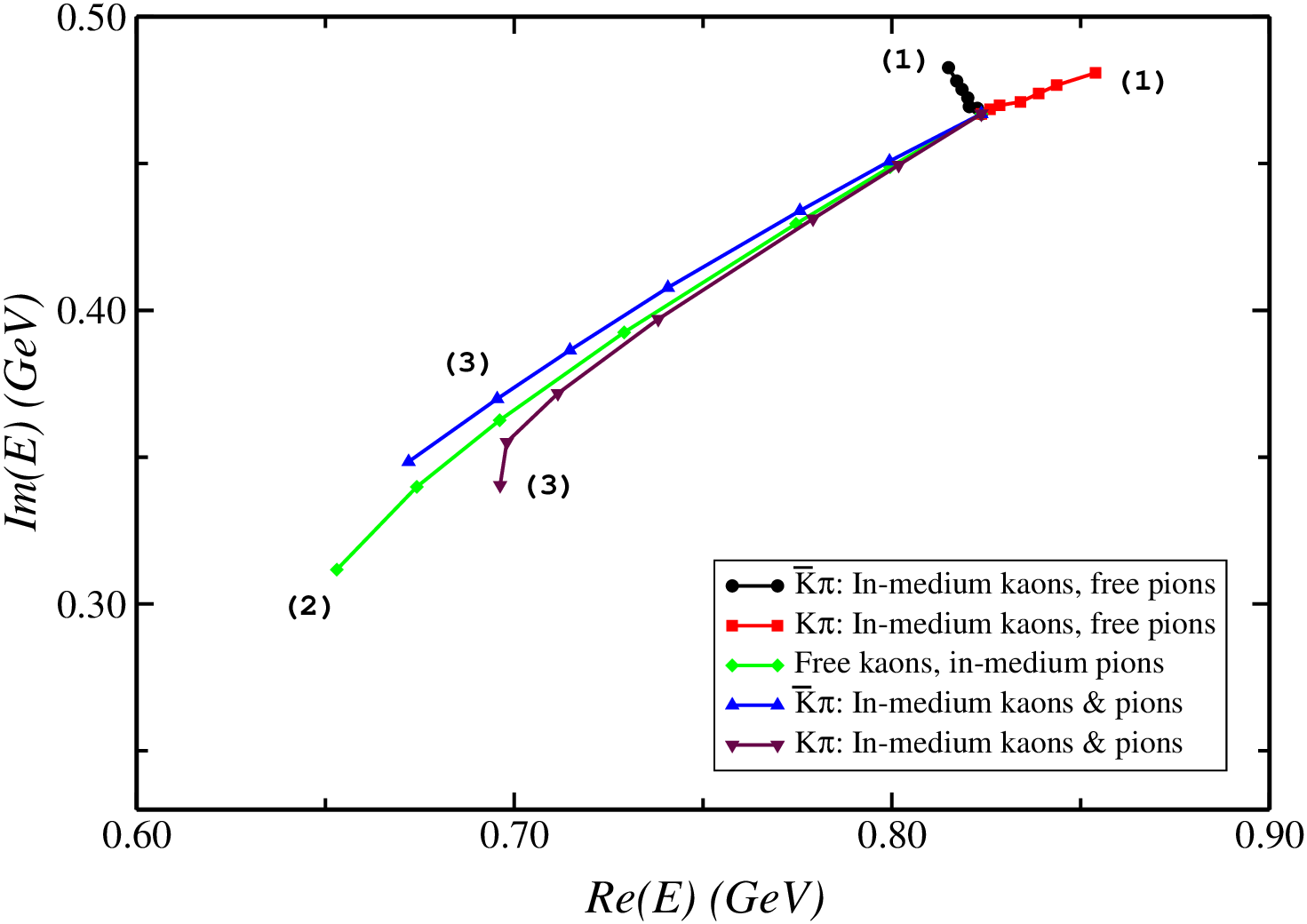}
\caption{\label{poles} (Color online) $\kappa$ pole trajectories at finite density. The
labels correspond to the three cases discussed in the text.}
\end{figure}

In Fig. \ref{poles} we show the $\kappa$ pole position in the complex energy
plane, using model A for the kaon self-energies. Every curve departs from a common
point, which corresponds to the pole position in the vacuum case, and we
increase nuclear density according to the following values: $\rho / \rho_0 =
0,1/8,1/4,1/2,3/4,1,3/2$, which correspond to the successive dots in each
trajectory. The curves labeled as '1' correspond to the case of in-medium
kaons and free pions. The first interesting fact is that the pole trajectory
splits up into two different branches corresponding to $K\pi$ and $\bar{K}\pi$
scattering. This was an expected result since the kaon self-energy is
asymmetric for the particle compared to the antiparticle. We can see that, in
the $K\pi$ branch, the $\kappa$ mass from the pole,
$M_{\kappa}=\textrm{Re}\sqrt{s}_{pole}$, moves to higher energies. This
responds to the repulsion felt by the kaons in the medium. In the $\bar{K}\pi$
branch one finds that the $\kappa$ pole mass decreases slowly with the nuclear
density. However, its decay width increases due to the opening of additional
decay channels in the medium, as $\kappa \to \pi M Y h$ and $\kappa \to \pi Y
h$. These channels are accounted for in the $\bar{K}$ $S-$ and $P-$wave
self-energies.

The next case that we consider corresponds to the curve labeled as '2' in Fig.
\ref{poles}, i.e., free kaons and in-medium pions. Since the pion self-energy is
the same for the three isospin components (in symmetric nuclear matter), we 
find a single trajectory for the $\kappa$ pole. When the nuclear density
increases, the pole position rapidly moves to lower energies.
This is consistent to what happens for the $\sigma$ meson and is due to the
strong attraction experienced by the pion in the nuclear medium.
 From this we
obtain that $M_{\kappa}$ is strongly reduced, from $825$ MeV in vacuum down to
$650$ MeV at $\rho=1.5\rho_0$, getting very close to the $K\pi$ threshold. 
$\Gamma_{\kappa}$ also shows a noticeable reduction of $300$ MeV at
$\rho=1.5\rho_0$. In fact, one may expect even a stronger reduction in the
$\kappa$ decay width, given the proximity of the pole to the $K \pi$ threshold and
the consequent reduction of available phase space for $\kappa \to K\pi$ decays.
However, this reduction of phase space is partly compensated by the simultaneous
opening of pion-related in-medium channels, namely $\kappa \to K\,
ph,\,K\,\Delta h$. Such effect was also discussed in
\cite{Chiang:1997di,VicenteVacas:2002se} for the $\sigma$ meson in $\pi\pi$
scattering.

Eventually, the two curves labeled as '3' in Fig. \ref{poles} represent the
$\kappa$ pole evolution with density in the full model A. We find again separate
$K\pi$ and $\bar{K}\pi$ branches, whose tendency to lower energies is a clear
signal of the strong influence of the attractive pion self-energy.
The $K\pi$ branch shows a sudden curvature for densities of and beyond $\rho_0$.

\begin{figure}
\includegraphics[width=0.7\textwidth]{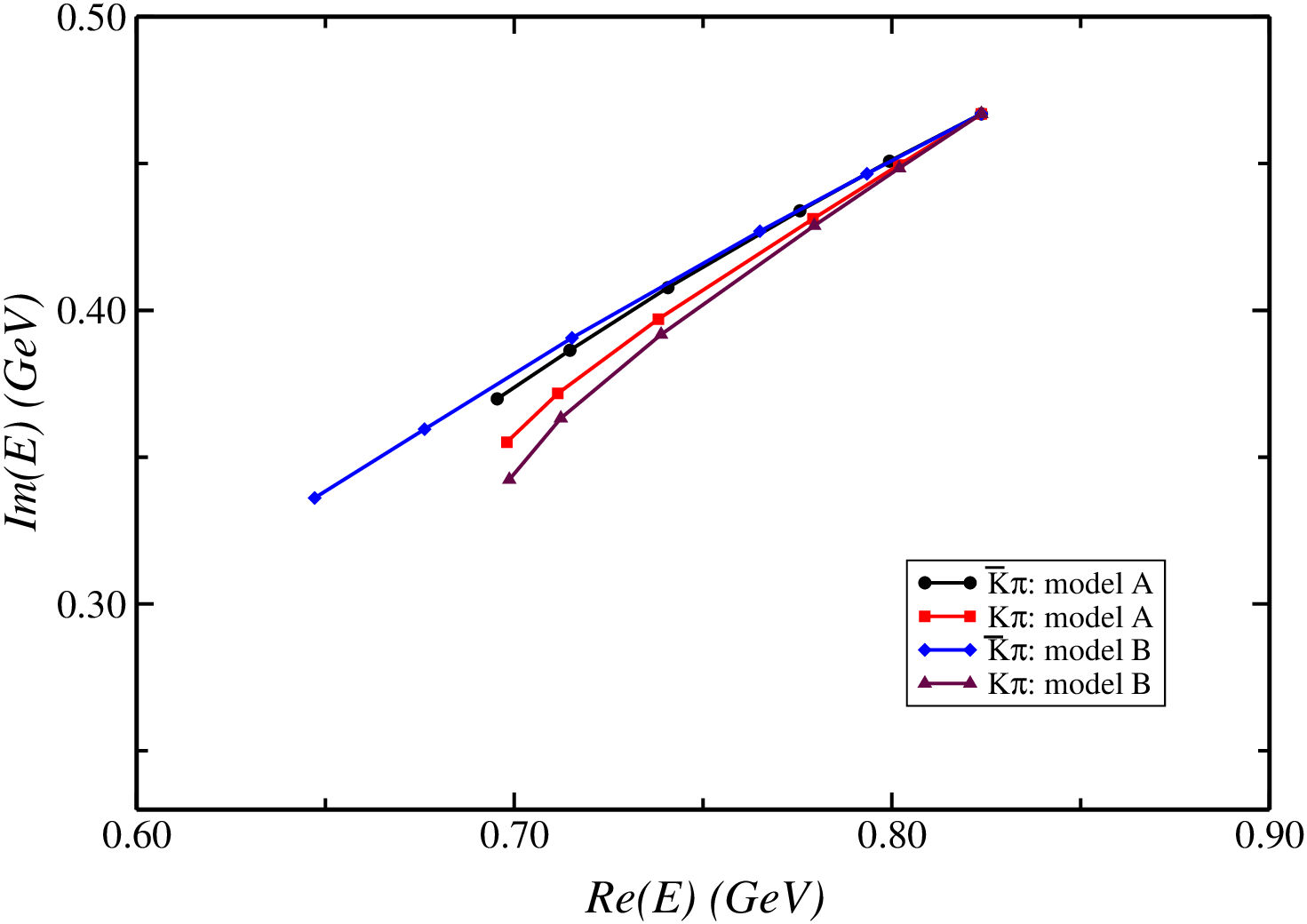}
\caption{\label{VLC_vs_Giessen} (Color online) $\kappa$ pole evolution with density. Comparison
between two models of anti-kaon self-energy.}
\end{figure}

In Fig. \ref{VLC_vs_Giessen} we compare the results on the $\kappa$ pole
position for models A and B and densities up to $\rho=\rho_0$. We observe that
the mass splitting between the $K\pi$ and $\bar{K}\pi$ branches is larger in
model B, amounting to about $50$ MeV at normal nuclear density. The
anti-$\kappa$ pole trajectories, sensitive to the $\bar{K}$ potential, show
stronger differences, whereas the $\kappa$ branches are quite similar since the
kaon self-energy only differs at the level of 10 \% in the two models.
Particularly, the $\bar{K}$ selfenergy in model B does not include an imaginary
part, leading to a further decrease of the $\kappa$ decay width.

\begin{figure}
\includegraphics[width=0.45\textwidth]{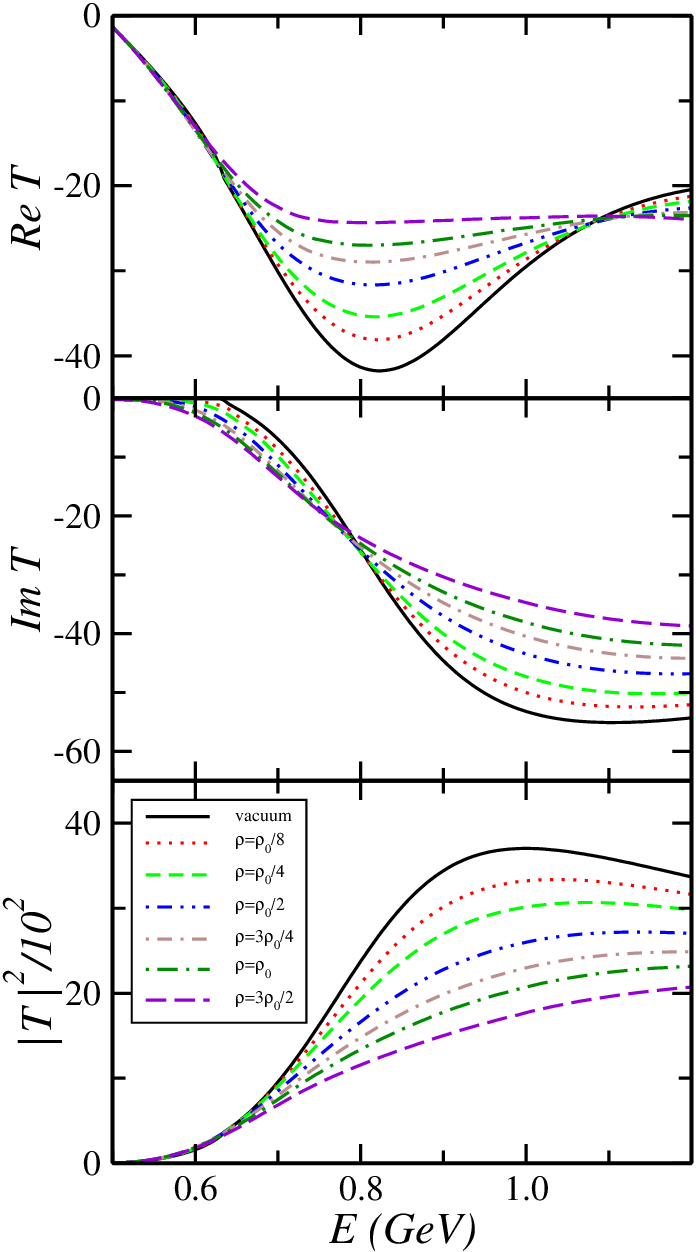}
\hspace{0.5cm}
\includegraphics[width=0.45\textwidth]{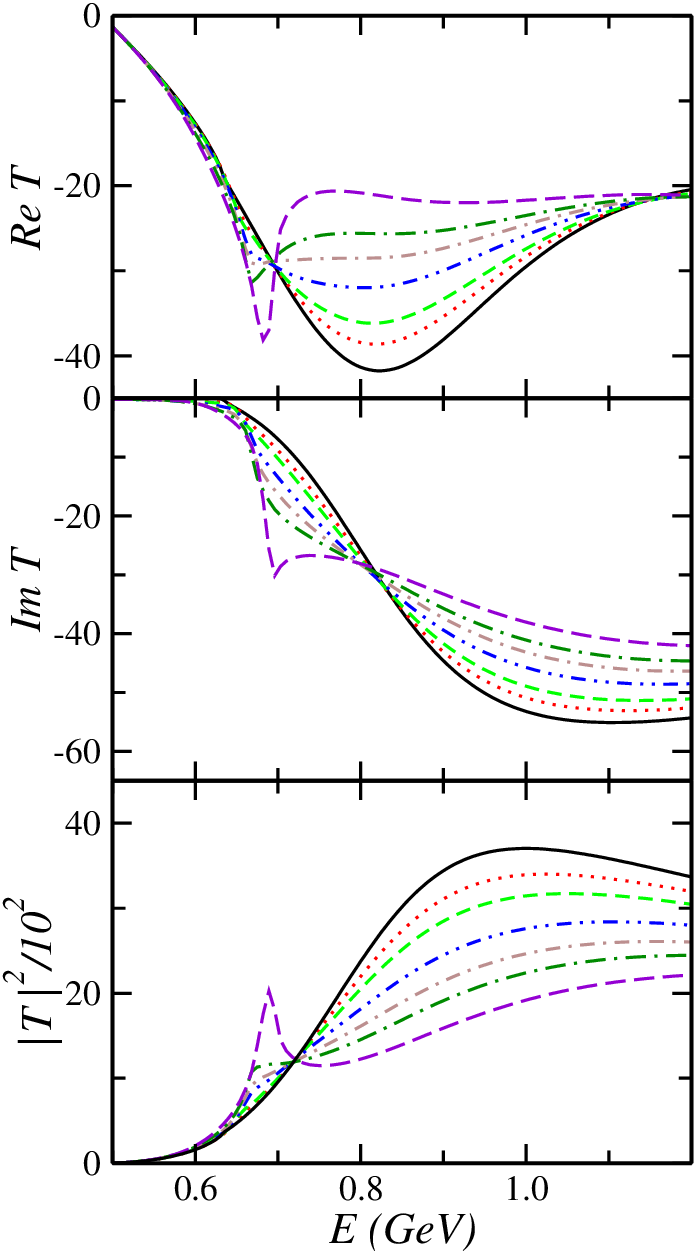}
\caption{\label{Tamplitudes} (Color online) Real, imaginary and squared modulus of the
kaon-pion scattering amplitude at several densities. Left panels correspond to
$\bar{K}\pi$ channel, right panels to $K\pi$ channel.}
\end{figure}

We have also calculated the $K\pi$ and $\bar{K}\pi$ scattering amplitudes for
several nuclear densities in model A. These amplitudes could be eventually
tested experimentally in reactions with pions and kaons in the final state.
The results are shown in Fig.
\ref{Tamplitudes}. The $\bar{K}\pi$ amplitude changes rather smoothly with
increasing density. The real part changes little at low energies and
flattens at higher energies. The imaginary part displays an increase of strength below $800$
MeV as compared to the vacuum case, and the threshold is shifted to lower
energies. Beyond $800$ MeV, though, we find a progressive decrease of strength.
The $K\pi$ channel shows a similar behaviour for $\sqrt{s} \gtrsim 800$ MeV as
the $\bar{K}\pi$ channel. However, we observe a greater accumulation of
strength below $800$ MeV in the imaginary part, which peaks at about $700$ MeV at
$\rho=1.5\rho_0$. The real part also reflects a rapidly changing structure in
this energy region. This effect was already observed in \cite{Chiang:1997di} for the
$\pi\pi$ scattering amplitude in the $\sigma$ channel, and it was found in
\cite{VicenteVacas:2002se} that this was correlated to a migration of the $\sigma$ pole to
lower energies at finite densities. A similar behavior is found here for the
$\kappa$ meson.
In the $\bar{K}\pi$ case some strength of $\textrm{Im}T$ spreads well below the vacuum 
threshold mainly because of the attraction experienced by the antikaon, that
moves the $\bar{K}\pi$ threshold down; and the coupling to in-medium channels with
the same quantum numbers occurring at lower energies, for instance meson
hyperon - hole excitations.
  However, for the $K\pi$ case the strength of $\textrm{Im}T$ is strongly accumulated above the
in-medium threshold, which is basically determined by the repulsive potential
affecting the kaon. The little strength below the vacuum threshold 
corresponds only to channels in which the pion is absorbed by a particle -
hole excitation.
We have also included in Fig. \ref{Tamplitudes} the squared modulus of the amplitude.
In the $\bar{K}\pi$ channel the main visible effect is a strong decrease of
strength beyond $700$ MeV, whereas in the region around and below threshold
little effect can be seen because of the lack of phase space, though it was
clearly visible in the imaginary part of the amplitude. The $K\pi$ channel clearly
exhibits a prominent structure close to threshold as commented above.

\begin{figure}
\includegraphics[width=0.9\textwidth]{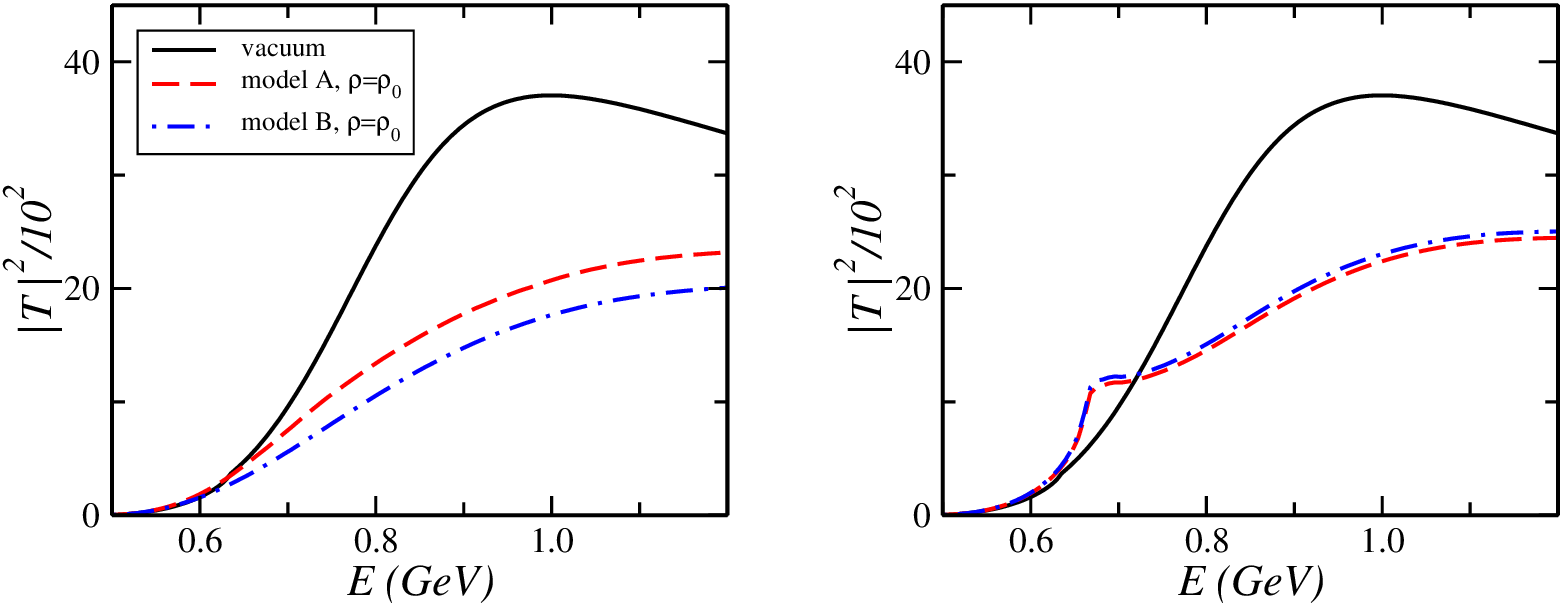}
\caption{\label{Tcompare} (Color online) Squared modulus of the kaon-pion scattering amplitude,
in vacuum and at normal density, 
for the two models of kaon selfenergy discussed in the text. Left: $\bar{K}\pi$
channel; right: $K\pi$ channel.}
\end{figure}

The resulting scattering amplitudes for model B show a similar trend as for
model A but there are some differences.  In Fig. \ref{Tcompare} we show the
squared modulus of the amplitudes for model B ,  at normal nuclear density,
together with model A and the free case for reference. The most visible effect,
as commented above, corresponds to the $K\pi$ channel where both models are
quite similar and there is little theoretical discussion concerning the $K$
potential in the medium. We find very similar results for this channel in the
two approaches. A stronger medium effect is observed in the
$\bar{K}\pi$ amplitude for model B, as a consequence of the more attractive
$\bar{K}$ potential in this model.

In summary, we find a noticeable modification of the $S-$wave $K\pi$
interaction in the $\kappa$ channel at finite densities. The $\bar{K}\pi$
channel is sensitive to the different $\bar{K}$ potentials used in this work.
The most visible medium effect, though, appears in the $K\pi$ channel, where the
$\bar{K}$ potential has little influence, and might be observed in
$K\pi$ invariant mass distributions around $650$ MeV.

\section{\label{s5}Conclusions}
We have studied the $\kappa$ meson properties in the nuclear medium. We follow
a chiral unitary model of $K\pi$ scattering that reproduces the elastic phase
shifts in the $I=1/2$ channel and dynamically generates the $\kappa$ meson,
which appears as a pole in the scattering amplitude. The medium effects have
been considered by dressing the pion and kaon propagators with selfenergies
that have been calculated elsewhere. When the density is increased, the $\kappa$
pole moves to lower masses and decay widths. However, it does not become as
narrow as to provide a clear signal to be observed experimentally. We have also
studied the scattering amplitude at finite densities for the $K\pi$ and
$\bar{K}\pi$ channels. We have found that the amplitude is strongly modified in
the medium in both channels, and particularly it is sensitive to the
differences between models of the anti-kaon potential. The most noticeable effect
is the accumulation of strength found at lower invariant masses of the $K\pi$
system. We suggest that this effect could be experimentally observed.

\begin{acknowledgments}
We acknowledge fruitful discussions with E. Oset. D.C. acknowledges financial
support from Ministerio de Ciencia y Tecnolog\'{\i}a. This work is
partly supported by DGICYT contract number BFM2003-00856 and the E.U.~EURIDICE
network contract no. HPRN-CT-2002-00311.
\end{acknowledgments}

\end{document}